\documentclass[journal]{vgtc}                

\ifpdf
  \pdfoutput=1\relax                   
  \usepackage{graphicx}                
  \DeclareGraphicsExtensions{.pdf,.png,.jpg,.jpeg} 
\else
  \usepackage{graphicx}                
  \DeclareGraphicsExtensions{.eps}     
\fi%

\graphicspath{{figures/}{pictures/}{images/}{./}} 

\usepackage{microtype}                 
\PassOptionsToPackage{warn}{textcomp}  
\usepackage{textcomp}                  
\usepackage{mathptmx}                  
\usepackage{times}                     
\usepackage{cite}                      
\usepackage{tabu}                      
\usepackage{booktabs}                  
\usepackage{bm}
\usepackage{graphicx}
\usepackage{xspace}
\usepackage{caption}
\usepackage[moderate]{savetrees}
\usepackage{balance}  

\DeclareRobustCommand*{\modern}{\fontfamily{cmss}\selectfont}
\DeclareRobustCommand{\etal}{\textit{et al.}\xspace}
\DeclareRobustCommand{\bfparhead}[1]{\noindent\textbf{#1}} 
\DeclareRobustCommand{\ctwo}{\textit{Clustrophile~2}\xspace} 

\onlineid{1014}

\vgtccategory{Research}

\title{Clustrophile 2: Guided Visual Clustering Analysis}

\author{Marco Cavallo and \c{C}a\u{g}atay Demiralp}
\authorfooter{
\item Marco Cavallo is with IBM Research. E-mail: mcavall@us.ibm.com.
\item \c{C}a\u{g}atay Demiralp is with MIT CSAIL \& Fitnescity Labs. E-mail:~cagatay@csail.mit.edu.
}

\shortauthortitle{Marco Cavallo and \c{C}a\u{g}atay Demiralp: Clustrophile 2}

\abstract{
 Data clustering is a common unsupervised learning method frequently used in
 exploratory data analysis. However, identifying relevant structures in
 unlabeled, high-dimensional data is nontrivial, requiring iterative
 experimentation with clustering parameters as well as data features and
 instances. The number of possible clusterings for a typical dataset is vast,
 and navigating in this vast space is also challenging. The absence of
 ground-truth labels makes it impossible to define an optimal solution, thus
 requiring user judgment to establish what can be considered a satisfiable
 clustering result. Data scientists need adequate interactive tools to
 effectively explore and navigate the large clustering space so as to improve
 the effectiveness of exploratory clustering analysis. We introduce  \ctwo, a
 new interactive tool for guided clustering analysis.  \ctwo  guides users in
 clustering-based exploratory analysis, adapts user feedback to improve user
 guidance, facilitates the interpretation of clusters, and helps quickly reason
 about differences between clusterings. To this end, \ctwo contributes a novel
 feature, the Clustering Tour, to help users choose clustering parameters and
 assess the quality of different clustering results in relation to current
 analysis goals and user expectations.  We evaluate \ctwo through a user study
 with 12 data scientists, who used our tool to explore and interpret
 sub-cohorts in a dataset of Parkinson's disease patients. Results suggest that
 \ctwo improves the speed and effectiveness of exploratory clustering analysis
 for both experts and non-experts.
}


\teaser{
  \centering
  \includegraphics[width=\linewidth]{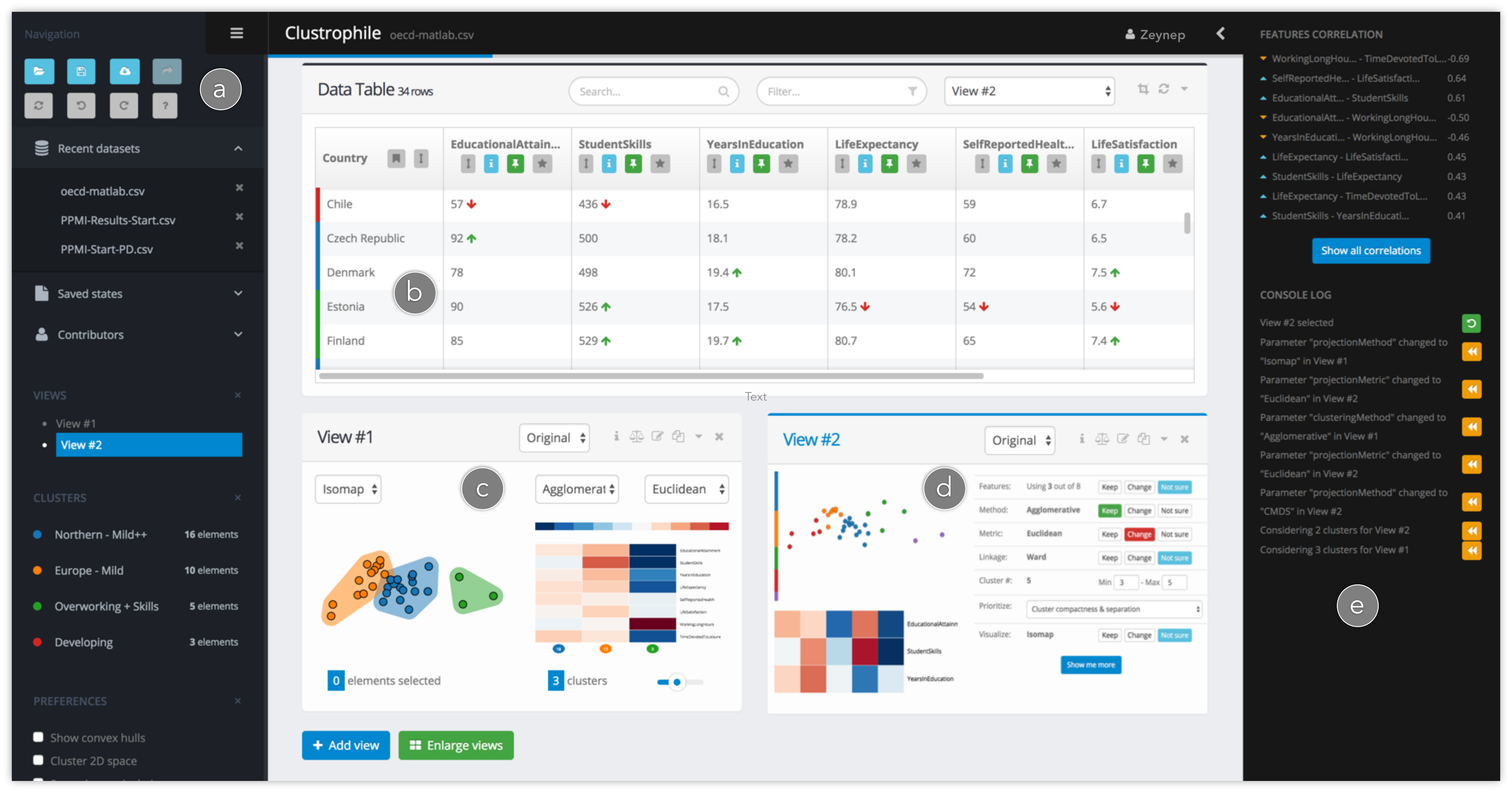}
  \caption{
  \ctwo is an interactive tool for guided exploratory
  clustering analysis. Its interface includes two collapsible sidebars (a, e)
  and a main view where users can perform operations on the data. \ctwo
  tightly couples b) a dynamic data table that supports a rich set of filtering 
  interactions and statistics and c) multiple resizable Clustering Views,
  which can be used to work simultaneously on different clustering instances. 
  Each Clustering View provides several ways to guide users during their analysis, 
  such as d) the Clustering Tour.
  \label{fig:teaser}} 
}

\renewcommand{\manuscriptnotetxt}{}

\vgtcinsertpkg

\begin{document}

\firstsection{Introduction}

\maketitle

The success of exploratory data analysis (EDA) depends on the discovery of
patterned relations and structures among data instances and attributes.
Clustering is a popular unsupervised learning method~\cite{Hastie_2005} used by
analysts during EDA to discover structures in data.  By automatically dividing
data into subsets based on a measure of similarity, clustering algorithms
provide a powerful means to explore structures and variations in data.
However, there is currently no single systematic way of performing exploratory
clustering analysis: data scientists iteratively combine clustering algorithms
with different data-transformation techniques such as data preprocessing,
feature selection and dimensionality reduction, and experiment with a large
number of parameters. This is an iterative process of trial and error based on
recurring formulation and validation of assumptions about the data. Data
scientists make multiple decisions in determining what constitutes a cluster,
including which clustering algorithm and similarity measure to use, which
samples and features (dimensions) to include, and what granularity (e.g.,
number of clusters) to look for.

The space of clusterings determined by different choices of algorithms,
parameters, and data samples and attributes is vast. The sheer size of this
exploration space is the first challenge in exploratory clustering analysis.
Data scientists need tools that facilitate iterative, rapid exploration of the
space of  data clusterings. The second important challenge is how to efficiently navigate this large space, rather than mere ad-hoc wandering. Therefore,
clustering tools would benefit from incorporating techniques that guide users,
imposing a structure over the clustering space that leads to efficient
navigation. Clustering is unsupervised by definition and we consider
here the most common case of complete absence of labels for validation (sometimes
referred to as ``fully unsupervised clustering''). 
If formal validation is not
possible, how do we estimate the meaningfulness of the outcome of a clustering
algorithm? Using the concepts of cluster compactness (closeness of data points
within the same cluster) and separation (how far a cluster is from others),
different \textit{internal} validation measures have been introduced to
estimate the ``goodness'' of a clustering and compare it to other clustering
results. Though widely used, these metrics fail to incorporate the context of
the analysis and the user's goals, prior knowledge, and expectations,
which often have significant role in determining the meaningfulness of a
clustering result.  Each internal validation metric takes into account slightly
different data characteristics and should be carefully chosen based on
the clustering task ~\cite{liu2010understanding}.  There is indeed no absolute
best criterion that, independently of the data and the underlying task, 
can establish the best result for the user's needs.


To address these challenges, we introduce \ctwo, a new interactive tool for guided clustering analysis.
\ctwo  guides users in clustering-based exploratory analysis, adapts user
feedback to improve user guidance, facilitates the interpretation of clusters,
and helps reason quickly about differences between clusterings. To this end,
\ctwo contributes a novel feature, the Clustering Tour, to help users choose
clustering parameters and reason about the quality of different clustering
results according to user's analysis goals and expectations.  We evaluate \ctwo
through a user study with 12 data scientists of varying skill sets who used our
tool to explore and interpret sub-cohorts in a dataset of Parkinson's disease
patients. We find that the Clustering Tour  enables data scientists to
utilize algorithms and parameters that they are unfamiliar with or hesitant to
use. Similarly, the Clustering Tour helps data scientists avoid prematurely
fixating on a particular set of data attributes or algorithmic parameters during
exploratory analysis.  In addition to the Parkinson dataset analyzed 
in the evaluation study, we use the
OECD Better Life (OECD for short) dataset~\cite{oecd} to demonstrate the use of
our tool in figures throughout the paper. The OECD dataset consists  of eight
numerical socioeconomic development indices of 34 OECD member countries. 

Below we first summarize related work and then discuss our design criteria for
\ctwo.  We then present the user interface of \ctwo along with integrated
visualizations and interactions, operationalizing the design criteria
presented. Then we introduce the Clustering Tour and the underlying graph
model.  Next we present a qualitative user study conducted with 12 data
scientists, followed by an in-depth discussion of the study results. We
conclude by summarizing our contributions and reflecting on lessons learned.

\section{Related Work}
\ctwo draws from prior work on interactive systems for visual clustering
analysis and guided exploratory data analysis. 

\subsection{Tools for Visual Clustering Analysis} 

Prior research applies visualization to improve user understanding of
clustering results across domains. Using coordinated visualizations with
drill-down/up capabilities is a typical approach in prior interactive tools.
The Hierarchical Clustering Explorer (HCE) ~\cite{Jinwook_Seo_2002} is an early
and comprehensive example of interactive visualization tools for exploring
clusterings. HCE supports the exploration of hierarchical clusterings of gene
expression datasets through dendrograms (hierarchical clustering trees) stacked
up with heatmap visualizations.  

Earlier research has also introduced clustering comparison techniques in
interactive systems~\cite{Lex_2010,Cao_2011,Lyi_2015,Pilhofer_2012,Jinwook_Seo_2002}.  DICON~\cite{Cao_2011} encodes statistical properties of clustering 
instances as icons and embeds them in a 2D plane through
multidimensional scaling. Pilhofer \etal~\cite{Pilhofer_2012} propose a method
for reordering categorical variables to align with each other, thus augmenting
the visual comparison of clusterings. Others have proposed similar visual
encoding techniques for comparing different clusterings of data dimensions with
applications to gene expression datasets in mind. To that end,
HCE~\cite{Jinwook_Seo_2002}, CComViz~\cite{zhou2009visually},
Matchmaker\cite{Lex_2010}, StratomeX~\cite{lex2012stratomex} and
XCluSim~\cite{Lyi_2015} all encode changes across clusterings of dimensions
essentially by tracing them with bundled lines or ribbons.

Researchers have also proposed tools that incorporate user feedback into
clustering. Matchmaker~\cite{Lex_2010} builds on techniques
from~\cite{Jinwook_Seo_2002} with the ability to modify clusterings by grouping
data dimensions. ClusterSculptor~\cite{Nam_2007} and Cluster 
Sculptor~\cite{Bruneau_2015} enable users to supervise clustering
processes. Schreck~\etal~\cite{Schreck_2009}
propose using user feedback to bootstrap the similarity evaluation in data
space (trajectories, in this case) and then apply the clustering algorithm.
FURBY\cite{streit2014furby} lets users refine or improve fuzzy clusterings
by choosing a threshold,  transforming fuzzy clusters into discrete ones. Sacha et al. \cite{sacha2018somflow} introduce a system for crime analysis where users can iteratively assign desired weights to data dimensions.

ClustVis~\cite{metsalu2015clustvis} uses both PCA and clustering heatmaps but
in isolation without interaction or coordination.
Clustrophile~\cite{demiralp2016clustrophile} coordinates heatmap visualizations
of discrete clusterings with scatterplot visualizations of dimensionality
reductions. It also enables correlation analysis and ANOVA-based significance
along with what-if analysis through direct manipulation on dimensionality
reduction scatterplots. Akin to Clustrophile,
ClusterVision~\cite{kwon2018clustervision} incorporates significance testing
and couples clustering visualizations with dimensionality reduction
scatterplots.  \ctwo extends Clustrophile with 1) new features to guide users
in clustering analysis, including the Clustering Tour, 2) a new task-driven
approach to improve cluster interpretation, and 3) broader and deeper support
for visual and statistical analysis, enabling the validation of multiple
clustering instances at a time. 

\subsection{Guiding Users in Exploratory Data Analysis} 

Earlier work in data analysis propose various tools and techniques to guide
users in exploring low-dimensional projections of data. For example, PRIM-9
(Picturing, Rotation, Isolation, and Masking---in up to 9
dimensions)~\cite{PRIM9_1974} enables the user to interactively rotate
multivariate data and view a continuously updated two-dimensional projection of
the data. To guide users in this process, Friedman and Tukey
~\cite{Friedman_1974} first propose the projection index, a measure for
quantifying the ``usefulness'' of a given projection plane (or line), and then
an optimization method, projection pursuit, to find a projection direction that
maximizes the projection index value.  The proposed index considers projections
that result in a large spread with high local density to be useful (e.g.,
highly separated clusters). In a complementary approach, Asimov introduces the
grand tour, a method for viewing multidimensional data via orthogonal
projections onto a sequence of two-dimensional planes~\cite{Asimov_1985}.
Asimov considers a set of criteria such as density, continuity, and uniformity
to select a sequence of projection planes from all possible projection planes.
%
%
Hullman \etal~\cite{hullman2013deeper} study how to generate visualization
sequences for narrative visualization,  modeling the sequence space with a
directed graph. Similar to Hullman \etal, \ctwo also models the visual
exploration space as a graph. However, \ctwo uses an undirected graph model and
focuses on modeling the clustering state space.
GraphScape~\cite{younghoon2017graphscape} extends Hullman \etal  with a
transition cost function defined between visualization specifications.  
While GraphScape purely considers chart specifications without
taking data or user preferences into consideration, \ctwo transitions consider
data, clustering parameters, and user preferences. 
 
Visualization recommender systems also model the visual exploration space
and evaluate various measures over the space to decide what to present the
user. For instance, Rank-by-Feature~\cite{seo:infovis04},
AutoVis~\cite{Wills_2008}, Voyager~\cite{Wongsuphasawat_2016}, 
SeeDB~\cite{Vartak_2015a}, and Foresight~\cite{Demiralp:2017:VLDB} use
statistical features and perceptual effectiveness to structure the presentation
of possible visualizations of data. \ctwo also provides methods for enumeration
and ranking of visual explorations. However, while recommendation systems
typically focus on suggesting individual charts based on attributes, \ctwo uses
the Clustering Tour to focus on clusterings and their visualizations,
complementing existing recommender systems.  SOMFlow~\cite{sacha2018somflow}
enables iterative clustering together with self-organizing maps (SOMs) to
analyze time series data. To guide users, SOMFlow also uses clustering quality
metrics. \ctwo goes beyond the use of quality metrics, considering user
feedback, clustering parameters, and data features along with 
interpretable explanations to guide users.

\section{Design Criteria}
We identify a set of high-level design criteria to be considered in developing
systems for interactive clustering analysis. These criteria extend the ones proposed in \cite{demiralp2016clustrophile} and are based on the regular feedback we received from data scientists during the development of Clustrophile and \ctwo.    


\bfparhead{D1: Show variation within clusters}
Clustering is useful for grouping data points based on similarity, thus
enabling users to discover salient structures.  The output of clustering
algorithms generally consists of a finite set of labels (clusters) to
which each data point belongs. 
In fuzzy clustering, the output is the
probability of belonging to one of those classes.  However, in both cases the
user receives little or no information about the differences among data points
in the same cluster. \ctwo combines complementary visualizations of the data---table,
scatterplots, matrix diagrams, distribution plots---to facilitate the
exploration of data points at different levels of granularity. In particular,
scatterplots represent dimensionally reduced data and thus provide a continuous
spatial view of similarities among high-dimensional data points.


\bfparhead{D2: Allow quick iteration over parameters} The outcome of a
clustering task is highly dependent on a set of parameters: some of them may be
chosen based on the type of data or the application domain, others are often
unknown a priori and require iterative experimentation to refine.  
\ctwo enables users to interactively update and apply
clustering and projection algorithms and parameters at any point while staying
in the context of their analysis session.


\bfparhead{D3: Promote multiscale exploration} 
The ability to interactively filter, drill down/up, and access details
on-demand is essential for effective data exploration.  \ctwo enables users to
drill down into individual clusters to identify subclusters as well as inspect
individual data points within clusters.   \ctwo lets users view statistical
summaries for each cluster and perform ``isolation''~\cite{Friedman_1974},
which enables splitting clusters characterized by mild features into more
significant subclusters.  Dynamic filtering and selection of single data points
are also implemented and coupled with statistical analysis to identify and
eventually remove outliers and skewed distributions in the data.

\bfparhead{D4: Represent clustering instances compactly} 
It is important for users to be able to examine different clustering instances
fluidly and independently without visual clutter or cognitive overload.  The
\ctwo interface employs the ``Clustering View'' element as the atomic component
representing a clustering instance, the outcome of a clustering algorithm using
a choice of parameters and data features and samples.  Clustering View  pairs
a projection scatterplot and a clustering heatmap using two complementary
visualizations.  A compact, self-descriptive representation is also useful for
visually comparing different clustering instances.  \ctwo lets users work
simultaneously on multiple Clustering Views, which can be freely organized by
users across the interface and help them keep track of how different choices of
features, algorithms and distance measures affect clustering results. 

\bfparhead{D5: Facilitate interpretable naming} 
How to attach meaning to the ``learned'' structures in clustering is an
important yet challenging problem. It is essential to facilitate the
meaningful naming and description of clusters and clustering instances.  
For each cluster computed,  \ctwo designates the cluster centroid 
as the cluster representative and assigns its identifier as 
the cluster name. \ctwo lets the user freely rename the cluster and its clustering instance according to her understanding of the data.


\bfparhead{D6: Support reasoning about clusters and clustering instances} 
Users often would like to know  what features (dimensions) of the data points
are important in determining a given clustering instance, or how different
choices of features or distance measures might affect the clustering, or
whether it is a ``good'' clustering.  Users also would like to
understand the characteristics of data points in a given cluster that
distinguish the cluster from other clusters and how these data points come to
be in the cluster. \ctwo dynamically chooses a combination of metrics based on
data and user preferences in supporting clustering analysis. It also includes
automated metric suggestions, visual explanations (e.g., decision-tree based
cluster visualization), quantitative indicators (e.g., stability and confidence
scores), and textual descriptions and hyperlinks to online references to help
user better interpret results and make informed decisions, eschewing the blind
use of clustering parameters and validation methods. 

\bfparhead{D7: Guide users in clustering analysis} 
Due to the number of possible combinations, iterative experimentation on
different clustering parameters can be non-trivial or time consuming, and
becomes even more challenging in a high-dimensional dataset.  Furthermore, most
users do not know in detail the advantages and disadvantages of clustering or
projection methods, sometimes choosing them blindly and simply trying all
possible parameter combinations.  It is thus important that the system provide
assistance to the user in navigating complex clustering spaces, while
incorporating the user's feedback in the process.  \ctwo provides textual
explanations with suggestions on when it could be worth using certain
parameters  with references (hyperlinks) to existing literature. \ctwo also
provides automated suggestions based on the dataset currently being analyzed, on
previous computations and on user preferences.  \ctwo introduces a novel
feature, the Clustering Tour. The Clustering Tour recommends a sequence
of clusterings based on clustering configuration choices, data features, and user
feedback. It samples the clustering space, promoting coverage in the absence of
user feedback. When the user ``likes'' a recommended clustering, \ctwo
recommends ``nearby'' clusterings. 

\bfparhead{D8: Support analysis of large datasets} The ability to
\textit{interactively} explore and analyze large datasets is important for
analysts in many domains and has been a major request of our collaborators.
\ctwo adopts caching, precomputation,
sampling and feature selection to support analysis with larger datasets. 
Addressing computational scalability also helps mitigate the visual scalability issues. 
\ctwo  also supports common interaction techniques such as panning \& zooming and visual grouping 
with smooth convex-hull patches to reduce visual clutter. 





\bfparhead{D9: Support reproducibility} One of
the primary motivations for data analysts in using interactive tools is to
increase productivity or save time. The iterative nature of a clustering
analysis continuously forces users to try out different parameters and
features, perform a set of computations, and decide which of the many directions
to take next---making the analysis session extremely hard to reproduce. \ctwo
logs  each operation performed by users, enabling them to undo/redo
single operations, to review the workflow of their analysis and to share 
it with their collaborators. 



\begin{figure}[t] 
  \centering 
  \includegraphics[width=0.5\textwidth]{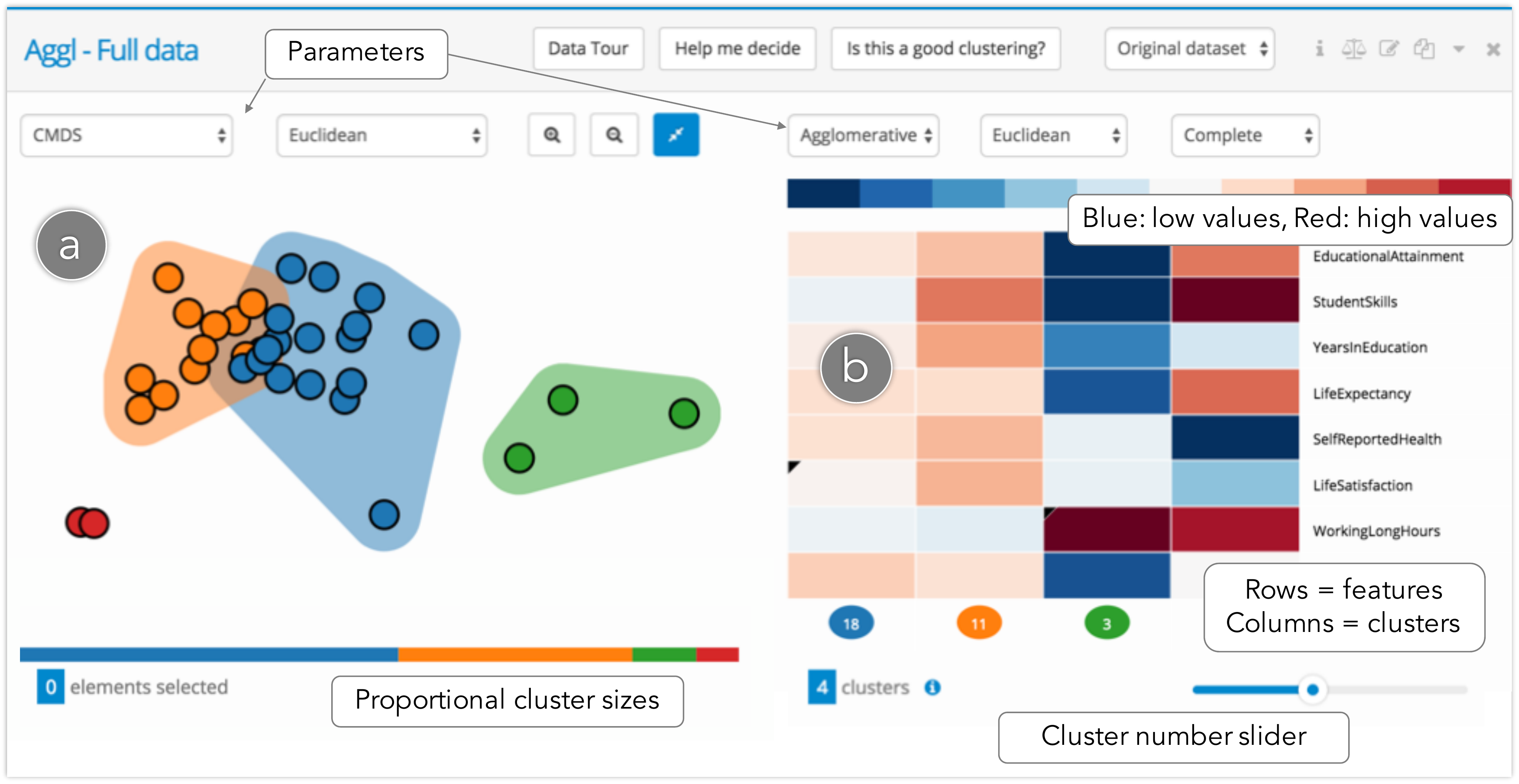}
  \caption{Clustering View, representing a clustering instance. a) A scatterplot shows the rows of the dataset projected on a 2D plane,
    where distance encodes the similarity between data points, whereas b) a
    heatmap allows easy comparison of clusters (represented by columns) by
    feature (row). \ctwo supports displaying multiple Clustering Views at a
    time, allowing users to compare different clustering results.
    \label{fig:view}
    \vspace{-0.5em}
  } 
  \end{figure}

\section{User Interface and Interactions}
In this section we briefly describe the main components of the \ctwo interface
and interactions. \ctwo has been developed iteratively according to the design
considerations introduced in the previous section. We refer back to the
relevant design criteria to motivate our design choices. The \ctwo interface
consists of  a main central view Fig.~\ref{fig:teaser}, two collapsible
sidebars (left Fig.~\ref{fig:teaser} and right Fig.~\ref{fig:teaser}) and
multiple modal windows.

The left sidebar (or Navigation Panel) contains a button menu to import
datasets from comma-separated-values (CSV) files, load data from previous
analyses and export the results (i.e. clusters, chart images) of the current
session. \ctwo supports saving the current state of the analysis (D9) for
follow-up analysis and sharing it with contributors who are also listed in the
Navigation Panel.
The right sidebar (hidden by default) logs operations and
parameter changes made by the user (Fig.~\ref{fig:teaser}e), enabling him to easily
revert the analysis to a previous state (D9). A convenient list of the top
pairwise feature correlations in the dataset is also displayed, facilitating a quick overview of statistical
dependencies.
The main view is subdivided into an upper region containing the Data Table
(Fig.~\ref{fig:teaser}b) and a lower region that displays one or more clustering
views (Fig.~\ref{fig:teaser}c,d). In fact, \ctwo enables data scientists to work simultaneously on multiple
clustering instances (D4), but at the same time links the coordinated Data
Table view to only one instance at a time. The currently selected clustering
instance is generally the one the user last interacted with, and its
corresponding Clustering View is marked with a blue header. The selected
instance and its cluster names are also made available in the Navigation Panel
(Fig.~\ref{fig:teaser}).

\begin{figure}[t] 
  \centering 
  \includegraphics[width=0.5\textwidth]{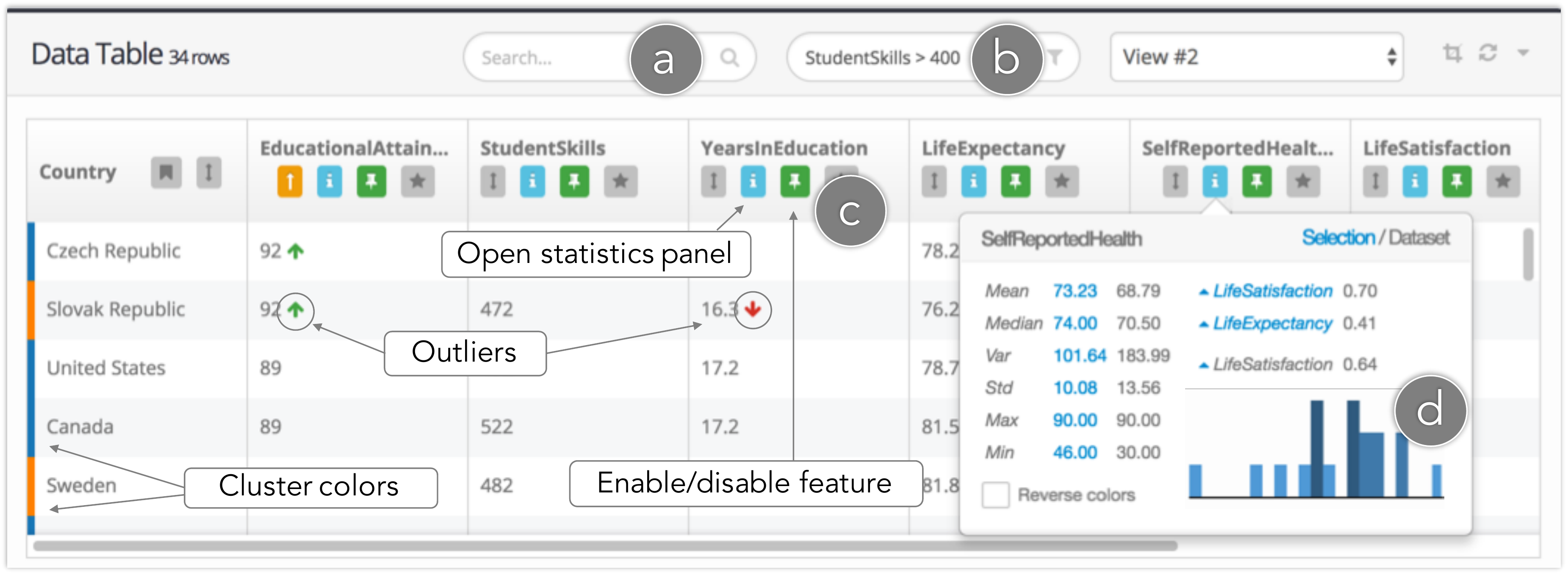}
  \caption{Data table showing the dataset associated with the current Clustering
    View in tabular format.  A user can dynamically filter the table by a)
    searching text and b) matching expressions. The user can also c) enable/disable
    data attributes to exclude/include them in the analysis, sort the table by
    a feature value, and display feature histograms along with summary
    statistics. For each data point, a color band  shows the assigned
    cluster and up and down arrows, respectively green and red, highlight
    high- and low-value outliers. d) The statistics panel  shows, for each
    feature, histogram density estimation, aggregate statistics, and top
    correlations for the current attribute with respect to the current data selection (blue) or the whole dataset (gray).\label{fig:datatable}} 
    \vspace{-1em}
  \end{figure}

\subsection{Visualization Views}
\bfparhead{Clustering View} A Clustering View (Fig.~\ref{fig:view}) represents
a single clustering instance and has the goal of both visualizing the
identified clusters and characterizing them based on their distinctive
features. In our user interface, the Clustering View also lets the user
dynamically change projection and clustering parameters for an associated
clustering instance, and keeps them always visible for easier comparison with
other Clustering Views.

The minimal set of visualizations we choose to summarize a clustering instance
consists of a scatterplot (Fig.~\ref{fig:view}, left) and a heatmap
(Fig.~\ref{fig:view}, right).
The scatterplot provides a two-dimensional projection of the data obtained
using dimensionality reduction and encodes clustering assignments through
color. Since clustering algorithms divide data into discrete groups based on
similarity, projections are a natural way to represent different degrees of
variation within and between groups as distance between elements (D1). Each
cluster of points can also be identified by a colored convex hull, simplifying
the visualization in cases with larger numbers of data points (D8).

The heatmap visualizes each cluster based on the aggregate feature values of the data points in the cluster. Each column of the matrix represents a cluster; rows
represent data features (dimensions). The color of each cell encodes the
average value of cluster members for a specific feature with respect to the
feature distribution. For instance, in the heatmap in Fig.~\ref{fig:view} the
dark red cell indicates that the Red cluster is characterized by very high
\textsc{WorkingLongHours}, whereas the dark blue cells indicate that the same
cluster has very low \textsc{EducationalAttainment} and \textsc{StudentSkills}
(i.e., red means higher values, blue lower values). This way, each cluster can
be quickly described (D6) by observing the heatmap vertically (e.g. intense
colors indicate the key features identifying a cluster, mild colors indicate
average values); similarly, clusters can be compared by looking horizontally at
the matrix diagram (e.g., from the second row of the heatmap, it is easy to see
that the green cluster is the one with highest \textsc{StudentSkills}). By
hovering on each cell, the user can inspect the average feature value of each
cluster and the p-value associated with the current selection feature algorithm
(which encodes the relevance of a feature).
Clusters are ordered from largest to smallest and display their member number
and color right beneath each column. Since with high-dimensional datasets (D8)
the number of rows would become too large, we display only the top relevant
features, which are chosen automatically by a feature selection algorithm (more
on this later) or manually selected by the user.

Users can select one or more data points or clusters from both the scatterplot
and the heatmap. When a selection is performed, it is reflected in both
visualizations and the Data Table. The isolation feature further lets users
re-cluster and re-project only the selected points, an operation particularly
useful for finding subclusters (D3).
From the Clustering View, users can dynamically change the parameters
associated to the associated clustering instance. Currently supported
clustering methods include K-means, Agglomerative (Hierarchical),
Spectral~\cite{shi2000normalized}, DBSCAN~\cite{ester1996density},
Birch~\cite{zhang1996birch}, Cure~\cite{guha1998cure} and
CLIQUE~\cite{agrawal1998automatic} algorithms that, as applicable, can be
combined with ten different clustering metrics and three types of linkage
strategies. Six types of projection methods are also available:
PCA~\cite{tipping1999probabilistic}, MDS~\cite{Kruskal_1964},
CMDS~\cite{torgerson:pm52}, t-SNE~\cite{maaten2008visualizing},
Isomap~\cite{tenenbaum2000global} and LLE~\cite{roweis2000nonlinear}. Users can
also define  custom projection and clustering algorithms and metrics. We note
that by default \ctwo applies dimensionality reduction and clustering in
high-dimensional space, and then visualizes the results using, respectively, a
scatterplot and a heatmap. 

The user can control the number of displayed clusters through a slider located
underneath the heatmap (Fig.~\ref{fig:view}). Different numbers of clusters are
automatically precomputed by \ctwo based on user settings, so that the user can
quickly change the number of clusters without waiting for further computations
(D8).  Another parameter that can be chosen from the clustering view is the
sampling rate of the data; this is useful for doing clustering in the presence
of larger datasets (D8).

\bfparhead{Data Table} While the Clustering View provides a high-level summary
of a clustering instance, it is fundamental for data scientists to be able to
drill down in the data and inspect individual data samples (D3). The Data Table view
gives the user the raw data, supporting statistical analysis, automatic outlier
detection, selection, and filtering. These features in particular make it
possible to reason about how single data points and feature distributions
affect the current clustering, and help the user decide which dimensions should
be considered or excluded by the clustering algorithm.

The Data Table (Figure~\ref{fig:teaser}b) contains a dynamic table
visualization of the current dataset in which each column represents a feature
(dimension) and each row represents a data sample.  The Data Table displays the
data and cluster assignments associated only with the currently selected
Clustering View. For each row, a vertical colored band encodes the cluster of
membership of the associated data sample,
whereas a set of green or red arrows respectively identify particularly high or
low feature values (``outliers'') with respect to each feature distribution
(Figure~\ref{fig:datatable}b).
Clicking on the buttons next to each feature name orders rows by cluster or by
column and displays basic statistics on a particular feature in a pop-up window
(Figure~\ref{fig:datatable}d).
In particular, \ctwo can compare the statistical values computed on the
currently selected rows and those of the whole dataset, plus displaying a histogram plot of the feature
distribution. A list of the features that correlate most to the selected
feature is also given, allowing quick discovery of data trends.
The search functionality (Figure~\ref{fig:datatable}a) lets users select data
samples using an arbitrary keyword search on feature names and values. Users can
also filter the table using expressions in a mini-language
(Figure~\ref{fig:datatable}b). For example, typing $age > 40$ $\&$ $weight < 180$
dynamically selects data points across visualizations in which the fields
{\modern age} and {\modern weight} satisfy the entered constraint. When some
rows are selected, the corresponding points of the scatterplot and cluster
columns in the heatmap in the current Clustering View are highlighted.


\begin{figure}[t] 
  \centering
  \includegraphics[width=1\columnwidth]{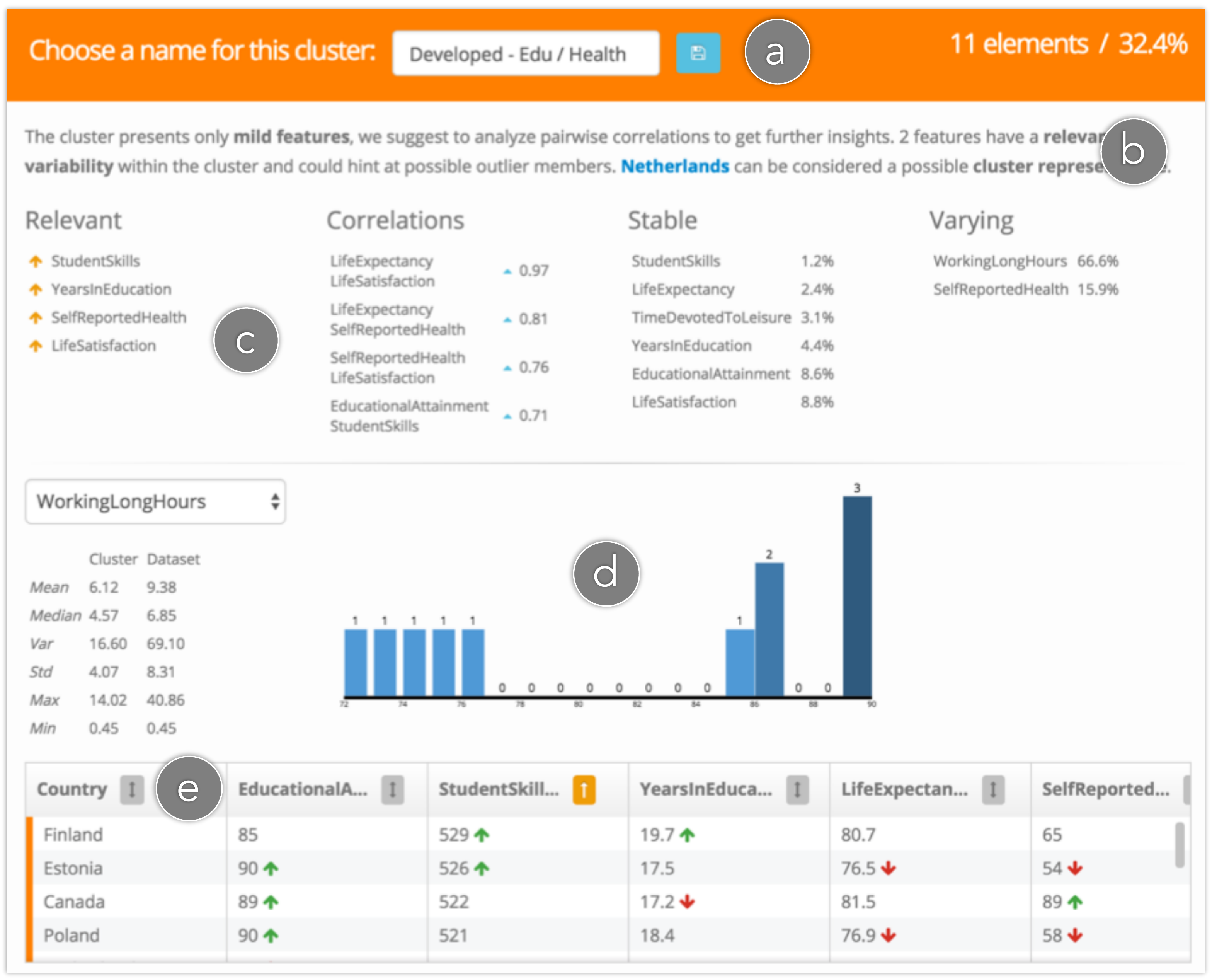} \caption{
    Cluster Details modal window. Double-clicking on a cluster lets the user access
    cluster-specific statistical information. The modal window shows b) an 
    automatically generated description for the cluster, to which the user can a) associate a custom name.
    The modal provides overall c) relevance and correlation analysis and  
    d) detailed aggregate statistics and a histogram density estimation of  
    the currently selected attribute for the data points in the cluster. 
    This modal view also includes e) an interactive a data table containing only the members of the cluster.
    \label{fig:cluster_details}
    \vspace{-0.5em}
  } 
\end{figure}

\bfparhead{Cluster Details} While the Data Table works well for inspecting
single data points and feature distributions across the dataset, the Cluster
Details modal (Fig.~\ref{fig:cluster_details}) aims at a deeper  
characterization of a specific cluster (D3). This modal can be opened by
double-clicking on any cluster in the user interface and contains statistical
information about the members of the cluster---such as most relevant features,
top pairwise feature correlations and outliers. The user can use this view to
assign a custom name to a cluster or to display the histogram for each feature
distribution with respect to the cluster. An automatically generated cluster
description containing suggestions for the analysis is also displayed.

\begin{figure*}[t] 
  \centering 
  \includegraphics[width=\textwidth]{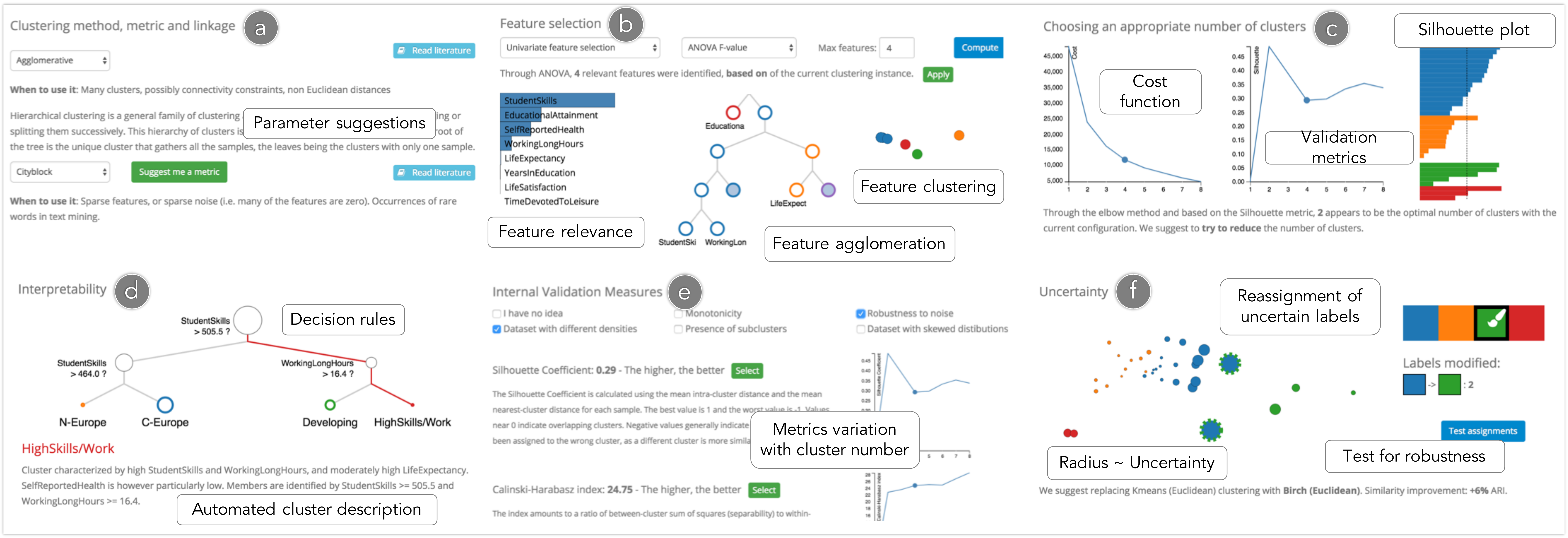}
  \caption{
    Providing guidance in clustering analysis. The figure displays only a
    subset of the views included in the ``Help me decide'' (top row) and the
    ``Is this a good clustering?'' (bottom row) panels of each Clustering View.
    a) Textual explanations and hyperlinks are used to suggest clustering
    parameters, b) different feature selection algorithms and visualizations
    are used to understand the relevance of data dimensions, and c) cost
    function and metric plots are used to suggest a good number of clusters. To
    evaluate the ``goodness'' of a clustering, d) decision rules and automated
    cluster descriptions are used to foster interpretability, e) several
    evaluation metrics are dynamically suggested, and f) uncertain clustering
    assignments are visualized and tested. \label{fig:guidance}} 
    \vspace{-0.5em}
\end{figure*}

\subsection{Raising Awareness About Choosing Parameters}

Given the high number of parameter combinations that may influence a clustering
outcome, it is important to guide users towards a reasonable choice of
parameters in the context of the current analysis. From each
Clustering View, the user can access a ``Help me decide'' panel
containing a tab dedicated to each parameter (D7).

\bfparhead{Feature selection} The choice of which features of the original
dataset to feed to the clustering algorithm can strongly influence both the
performance and the quality of clustering results. To help the user understand
if and which data dimensions should be included in the analysis, \ctwo provides
a list of the most relevant data dimensions according to several
feature-selection algorithms useful in removing features with low variance or
high pairwise correlation (e.g. variance threshold, feature agglomeration,
random projection) and in filtering out noisy or uninfluential features (e.g.
univariate feature selection based on chi-squared or ANOVA f-value, and recursive
feature elimination~\cite{guyon2002gene}).  \ctwo also introduces a
hierarchical clustering of the dataset's features (Fig.~\ref{fig:guidance}b),
displaying, through a scatterplot and a dendrogram, how data dimensions can be
grouped together based on similarity (feature agglomeration). 

\bfparhead{Sampling} In presence of larger datasets (more than 10,000 data
samples) \ctwo suggests that the user perform clustering only after sampling
the original data, in order to speed up the computation during the initial data exploration phase. Since the analysis is
unsupervised, the user can only either change the percentage of random sampling, or disable sampling when he wants to validate his findings on the whole dataset.

\bfparhead{Clustering algorithm, metric and linkage} For each possible choice
of clustering parameters, \ctwo provides a textual description with theoretical
advantages/drawbacks and use cases for each method (Fig.~\ref{fig:guidance}a).
For instance, users can learn that Kmeans is not suited in the presence of
uneven cluster sizes and non-flat geometries, or that the Cityblock affinity can
outperform Euclidean distances in cases with sparse data. For
clustering metrics and linkage criteria, \ctwo can suggest to the user which
parameters to use by testing them asynchronously and picking the one that
generates the best cluster separation. Hyperlinks to related literature are
also included.

\bfparhead{Number of clusters} Clustering algorithms do not generally output a
unique number of clusters, since this is generally a user-defined parameter. By
generalizing the idea of the ``elbow plot'' for the K-means cost function,
\ctwo precomputes a range of clustering solutions, each with a different number
of clusters in a user-defined range, and compares them in a line chart
(Fig.~\ref{fig:guidance}c). In particular, the horizontal axis corresponds to
the number of clusters and the vertical axis represents the value of one of the
internal validation measures. Based on the metric formulation, the optimal
number of clusters is given by the maximum, minimum or elbow value of the line
chart \cite{liu2010understanding}. When applicable, \ctwo complements the line
chart with a clustering-algorithm-specific plot (e.g., a dendrogram for
hierarchical clustering). A silhouette plot~\cite{Silhouette} is also included
(Fig.~\ref{fig:guidance}c, right), providing more detailed information on
which clusters should be merged and which data points are critical to
determining the optimal cluster number.

\bfparhead{Projection} Although the dimensionality-reduction method used to
visualize the scatterplot does not influence clustering results, it may
visually bias how a user perceives the quality of a given 
clustering instance.
To handle this, \ctwo provides a description and a set of references for each
projection method in addition to an automated suggestion. By precomputing each
projection, our tool applies to dimensionally reduced data the same internal evaluation metrics used for
clustering and suggests a  projection
algorithm that optimizes cluster compactness and separation in the scatterplot.

\subsection{Guiding Users Towards a Better Clustering} Once clustering
parameters are chosen, the next step is assessing the quality of a clustering
outcome. In the panel ``Is this a good clustering?'', \ctwo aims at helping the
user reason about the absolute and relative ``satisfactoriness'' of the results (D6).

\bfparhead{Quantitative validation} Since no ground truth labels are available,
internal validation measures are the only objective numerical values for
assessing the goodness of a clustering instance and comparing it to other
instances. Instead of adopting only one metric, \ctwo acknowledges the pros and
cons of each measure and tries to help the user choose the measure that better
fits the data and requirements. Using Liu et al.'s work
\cite{liu2010understanding}, we associate the performance of each validation
metric with a set of five conditions: presence of skewed distributions,
subclusters and different cluster densities; robustness of the algorithm to
noise; and monotonicity of the measure's cost function. While the first three
can be automatically inferred from the data, the last two are dictated by user
preferences. For instance, using Silhouette in the presence of subclusters or
using Calinski-Harabasz with noisy data could lead the user to a non-optimal cluster
choice.  On top of briefly describing each measure, \ctwo filters measures
dynamically, showing only those that match the user's interests and displaying how
their values change based on the number of clusters (Fig.~\ref{fig:guidance}e).

\bfparhead{Interpretability} We believe that, in addition to a quantitative
evaluation, a qualitative analysis of the clustering results is fundamental in
understanding if the user's goal has been reached.
Even if the user is exploring the data freely, it is important to interpret each cluster in relation to its features. To this end, we
apply decision trees~\cite{breiman2017classification}, in combination with
cluster average feature values, as an approximate and generalizable solution
for cluster interpretability. Once a clustering result is obtained, we use its
clustering assignments as ground-truth labels to train a decision-tree
classifier, whose decision rules are then displayed in a tree diagram
(Fig.~\ref{fig:guidance}d). By interactively exploring the decision
tree, the user can reason about the main features used to distinguish data
points in different clusters. By combining decision tree paths and the information
presented in the Clustering View's heatmap, \ctwo also provides a textual
description of each identified cluster.

\begin{figure}[ht] \centering
  \includegraphics[width=\columnwidth]{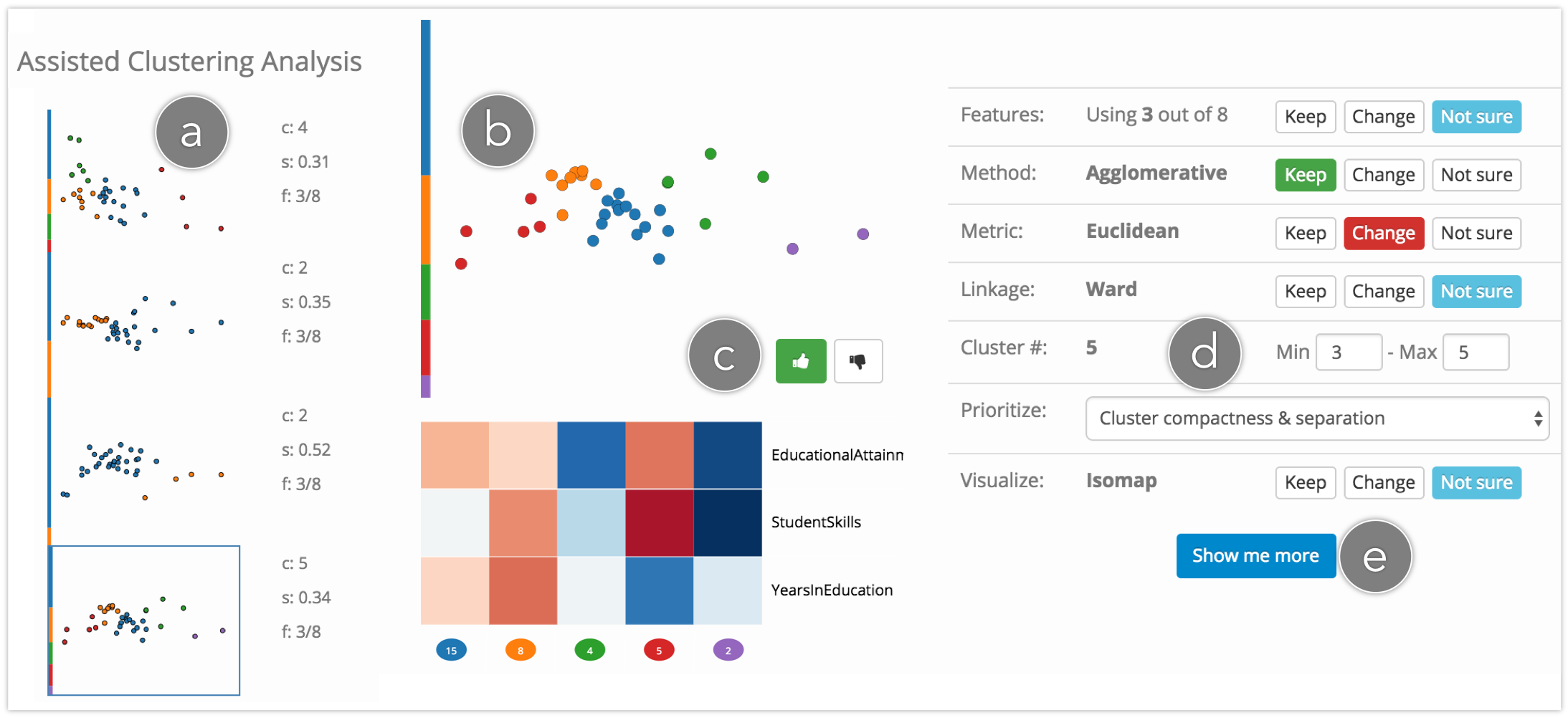} 
  \caption{Clustering Tour interface. Users can explore the possible clustering 
    solutions generated by our algorithm by simply clicking on the ``Generate solution'' 
    button. On the left, a) previous solutions are listed and compared, while b) the 
    current one is represented through a scatterplot and a heatmap visualization in the
    middle. Users can also define c) constraints on clustering parameters ,
  specifying which ones can be modified by the Clustering Tour algorithm.
\label{fig:datatour_interface}} 
\vspace{-0em}
\end{figure}

\begin{figure}[ht] 
  \centering 
  \includegraphics[width=\columnwidth]{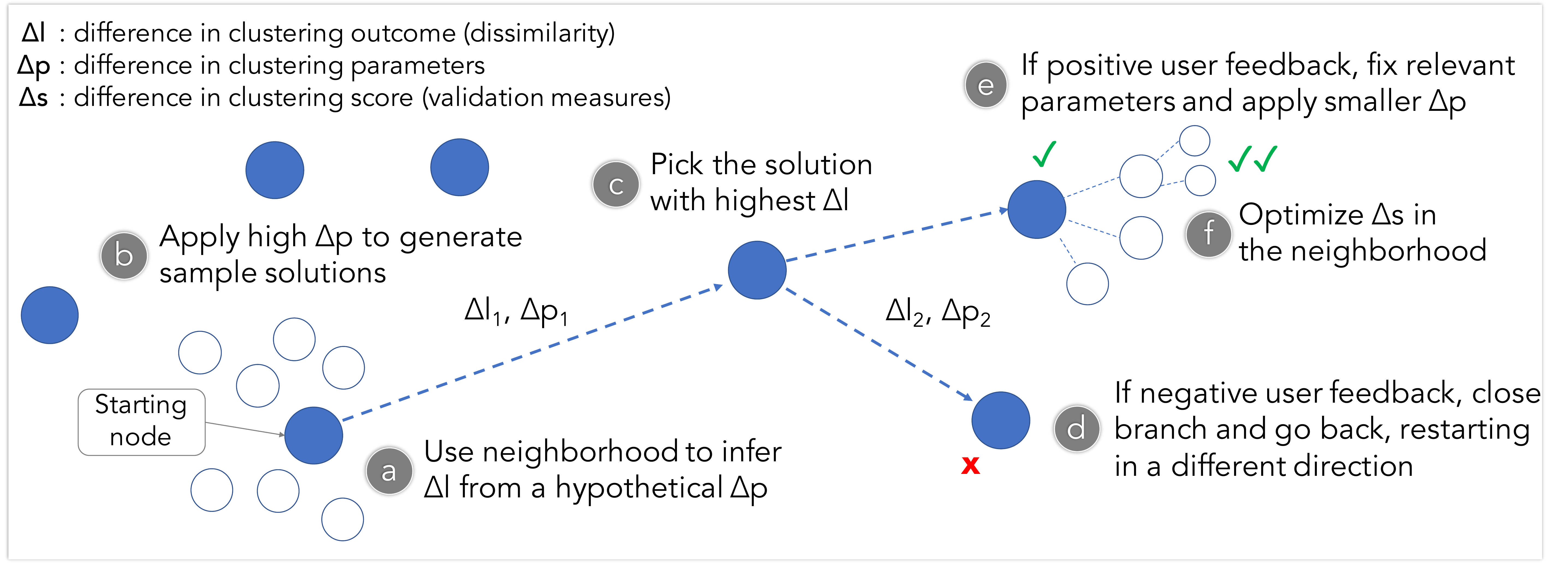}
  \caption{
    Clustering Tour model. In a graph of possible clustering solutions
    (nodes), we first explore outcomes that are very different from each
    other ($\Delta l$ encodes the distance between nodes). To this end, we
    estimate the impact of each clustering parameter ($\Delta p$) and sample a
    set of possible solutions that should optimally be distant from the current
    node. Once a user gives positive feedback on a clustering result, the
    neighborhood of the node is explored, applying smaller parameter changes
    and optimizing cluster separation ($\Delta s$).\label{fig:datatour_model}
  }
  \vspace{-1em}
\end{figure}

\begin{figure*}[t] 
  \centering
  \includegraphics[width=\textwidth]{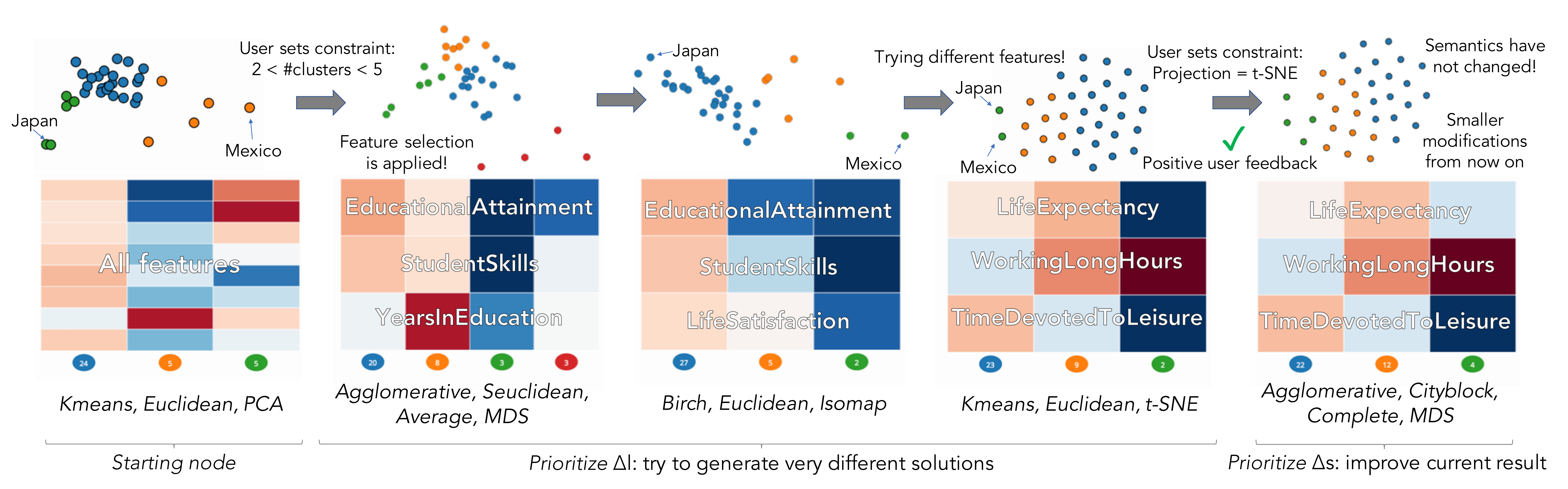} 
  \caption{Clustering Tour sequences for the OECD dataset. Starting from the
    initial clustering instance, the tour explores different parameter
    combinations, proposing alternative solutions to the user. In particular,
    by observing changes in the heatmap, we can interpret how the algorithm
    produces new unexpected and semantically interesting solutions. For
    instance, it is interesting to discover how very different countries such
    as Japan and Mexico can even cluster together. Once the user gives positive
    feedback, the tour proposes alternatives that are semantically similar to
    the selected one.\label{fig:datatour_example}
  \vspace{-1em}} 
\end{figure*}

\bfparhead{Uncertainty} Most clustering algorithms output discrete classes of
objects: either a data point belongs to cluster $A$ or it belongs to cluster
$B$. However, as can be seen in the Clustering View scatterplot, the position
of some ``outlier'' data points may suggest a forced clustering assignment.
Small differences in clustering parameters can easily cause uncertain data
points to change their cluster assignment, unbalancing the size of the clusters
involved and sometimes hiding some of their distinctive features. For this
reason we believe that being aware of low-confidence clustering assignments is
important, and we propose a dedicated panel where these critical points are
displayed through a variation of the Clustering View scatterplot
(Fig.~\ref{fig:guidance}f). When fuzzy clustering confidence values are not
available, we use the distribution of per-point silhouette scores to determine
which data points are uncertain. In particular, here we let users reassign
class for these points and find the combination of parameters that produces the
cluster assignment closest to the their expectations. This is currently done by
applying all combinations of clustering algorithms and metrics and by ranking
the outcomes based on their Adjusted Mutual Information 
score~\cite{vinh2010information}.

\section{Clustering Tour}
By iteratively changing all clustering parameters, a user can dynamically
explore the space of possible clustering solutions until a satisfactory
solution or set of insights on the data is found. However, even with guidance
in parameter choice, the space of possible parameter combinations and
clustering solutions is too large to explore manually. There are certain
parameter choices that affect the clustering outcome more than 
others. Overall, it is useful
to let users first explore the parameter choices determining solutions that are very
different from each other, metaphorically making large leaps across the space of
possible clusterings to enable a quick tour of the data. If the user
likes a solution and wants to refine it, then other parameter choices can be
made to explore solutions similar to the selected one. With this concept in
mind, we introduce a \textit{Clustering Tour} feature to help the user quickly
explore the space of possible clustering outcomes. The interface shown in
Fig.~\ref{fig:datatour_interface} contains (a) a list of previously explored
solutions, (b) a scatterplot and a heatmap representing the current solution,
(c) a set of buttons for the user to give feedback, and (d) a choice of
modalities by which the user can constrain how parameters are updated. 

\subsection{Navigating the Clustering Solution Space}

\bfparhead{Clustering descriptors}
To determine which clustering solutions (instances) to recommend to the user,
our algorithm considers three clustering descriptors: \textit{parameters}, 
\textit{labels}, and \textit{score}. Each combination of
parameters $p$ (including input features, clustering algorithm,
similarity measure and cluster number) generates a clustering outcome, which
consists of an array of assigned class labels $l$. Through a linear combination
of existing internal clustering validation measures, we assign a score $s$ for
a given clustering instance. If we change some of the clustering parameters by
a hypothetical amount $\Delta p$, we obtain a second clustering solution
whose class assignments differ from the previous ones by $\Delta l$ and whose
score differs by $\Delta s$. $\Delta l$, which we compute as $1 - AMI$ (Adjusted Mutual Information score~\cite{vinh2010information}),
aims to encode how much the two clustering instances (solutions) are
\textit{semantically} different from each other. $\Delta s$ is instead an
indicator of how much the second outcome generates clusters that are more
compact and better separated.

\bfparhead{The graph model}
We model the space of possible clustering solutions as an undirected graph,
where each solution is a node (Fig.~\ref{fig:datatour_model}). Since we want to
prioritize first the exploration of clusterings that have different outcomes
(i.e., prioritizing coverage), we set the distance between two nodes to be
$\Delta l$, so that solutions with similar clustering assignments lie close to
each other in the graph. Each edge between two nodes is also associated with
the difference in parameters $\Delta p$ and score $\Delta s$ between 
the nodes. 

\bfparhead{Bootstrap}
We consider the active clustering instance in the Clustering View to be the
entry node of the graph. From here, we want to apply a set of parameter changes
$\Delta p$ that would generate a significant modification $\Delta l$ in cluster
assignments. Since computing the full graph is not feasible, our strategy consists of sampling a set of nodes that are distant
from the current one by inferring $\Delta l$ from a hypothetical $\Delta p$. In
other words, we want to roughly estimate which parameter changes will create
the largest modification in cluster assignments.  To achieve this, we assign a
weight $w_i$ to each clustering parameter $i$ and compute a numerical
representation of $\Delta p$ as $\sum_{i} w_i * c_i$, where $c_i$ is the amount
of change for parameter $i$ (e.g., difference in number of clusters; defaults
to one for changes in algorithm). Weights are asynchronously estimated based on
the average $\Delta l$ produced by separately applying a randomized subset of
the admissible values for each parameter to the current clustering. For
instance, if modifying the number of clusters in the current solution produces
on average a $\Delta s$ higher than changing the clustering metric, cluster
number will have a higher weight in determining $\Delta p$.


\bfparhead{Updating parameters} Once parameter weights are computed, we sample
a set of possible $\Delta p$ by prioritizing changes in parameters with higher
weight.  In the absence of user-specified constraints, a set of alternative values
for each parameter is picked through random sampling and, among the generated
solutions, the one producing the highest $\Delta l$ is chosen. 
We select the data features (dimensions) to be considered 
for clustering by cyclically applying the feature selection methods
described in Section 4.2 and selecting the features with highest relevance
and/or lowest pairwise correlation. At the same time, we randomly exclude from
the analysis features ranked highly relevant to prevent single features from
biasing the clustering result. Once a clustering suggestion is established, we
perform a subset of the dimensionality reduction methods in Section 4.1 and
apply clustering validation measures on their output to choose the one that
best visualizes the separation among clusters.  

\bfparhead{User feedback}
Both the clustering result and the dimensionality reduction are shown to the
user (Fig.~\ref{fig:datatour_interface}), who can continue exploring different
solutions by pressing the ``Generate new solution'' button. When the user is
relatively satisfied with the current solution, he can press the ``I like it''
button to explore the neighborhood of the current node in the graph. In this
situation high-weight parameters (often features and cluster number) tend to
remain fixed, and lower-weight parameters (e.g., typically the clustering
metric) are changed to produce slight variations in the clustering outcome
(small $\Delta l$). 
Only at this stage validation measures are incorporated in generating 
clustering recommendations. Among the alternative solutions, now the one
that generates the highest $\Delta s$ is chosen. If the user presses the ``Very
bad :('' button, the Clustering Tour goes back to the previous node of the
graph and explores a different direction (i.e. tries to generate a solution
with high $\Delta l$ from the disliked solution). At any point in the
Clustering Tour, the user can constrain the variability of available
parameters, deciding which ones should be fixed or changed and which ones
should be decided by the algorithm. When the user is satisfied, he can decide
to apply the identified parameters to the associated Clustering View.

We illustrate in Fig.~\ref{fig:datatour_example} a sample execution of the
Clustering Tour on the OECD dataset, showing the results generated by the
algorithm based on user feedback.

\begin{table*}[t]
\resizebox{\textwidth}{!}{
\centering
\begin{tabular}{clllllllllllllllll}
  \toprule
  \textbf{Id} & 
  \multicolumn{1}{c}{\textbf{Archetype}} & 
  \multicolumn{1}{c}{\textbf{Domain}} & 
  \multicolumn{1}{c}{\textbf{Time}} & 
  \multicolumn{1}{c}{\textbf{Isolation based on}} & 
  \multicolumn{1}{c}{\textbf{Features}} & 
  \multicolumn{1}{c}{\textbf{Feat. Selection}} & 
  \multicolumn{1}{c}{\textbf{Algorithm}} & 
  \multicolumn{1}{c}{\textbf{Metric}} & 
  \multicolumn{1}{c}{\textbf{Projection}} & 
  \multicolumn{1}{c}{\textbf{Clusters}} & 
  \multicolumn{1}{c}{\textbf{Cluster Names}} & 
  \multicolumn{1}{c}{\textbf{Validation}} & 
  \multicolumn{1}{c}{\textbf{Tour}} \\
  \midrule 
1 & Hacker & No & 69m & - & 17& Custom & Agglomerative (Complete) & Euclidean & Isomap & 5 & 
  \begin{tabular}[c]{@{}l@{}} Mild, Tremor-dominant, Rigid, Hand \&\\ 
    feet mobility, Posture \& gait issues
  \end{tabular} & SDbw & No \\
2 & Hacker & No & 51m & Disease severity & 16 & Custom & Agglomerative (Average) & Cityblock & MDS & 5 & 
  \begin{tabular}[c]{@{}l@{}}Tremor at rest, Hand Tasks, Rigidity and\\ 
    expression, Posture and gait, Limbs agility
  \end{tabular} & Silhouette & No \\
3 & Hacker & Yes & 55m & -& 32 & Custom & Kmeans & Euclidean & t-SNE & 4 & 
\begin{tabular}[c]{@{}l@{}}Bilateral, Unilateral left, Unilateral\\ 
  right, Axial
\end{tabular} & SDbw & No  \\
4  & Hacker & Yes & 46m & 
\begin{tabular}[c]{@{}l@{}}Drug use, \\
  Outlier removal 
\end{tabular} & 33 & Custom & CLIQUE & Euclidean & LLE & 3 & Bradykinesia, Tremor, Dyskinesia                                          & - & Yes  \\
5  & Scripter & No & 39m & 
\begin{tabular}[c]{@{}l@{}}
  Affected side\\ 
  \& disease severity
\end{tabular} & 36 & ANOVA & Kmeans & Euclidean & PCA & 4 & 
\begin{tabular}[c]{@{}l@{}}Mild, Hand movements, Tremor,\\ 
  Rigid, (Arising from chair)\end{tabular} & - & No \\
6  & Scripter & No & 40m & - & 15 & PCA & Birch & Euclidean & MDS & 5 & Mild, Rigid, Tremor, Posture \& Gait, Extreme                  & - & Yes  \\
7  & Scripter & Yes & 62m & 
\begin{tabular}[c]{@{}l@{}}
  Drug use, Outlier\\ 
  removal, Random samples
\end{tabular} & 33 & PCA & Kmeans & Euclidean & PCA & 4 & Bradykinetic, Tremoring, Gait, Fine movements                                & - & No  \\
8  & Scripter & Yes & 24m  & Drug use & 13 & Custom & Kmeans & Euclidean & CMDS & 4 & 
\begin{tabular}[c]{@{}l@{}}
  Not impaired, Left, Right, Task driven \\ 
  impairment
\end{tabular}                                   & -                              & No  \\
9  & Application user  & No & 34m & - & 20 & Custom & Agglomerative (Ward) & Euclidean  & t-SNE & 5 & 
\begin{tabular}[c]{@{}l@{}}Mild, Limbs agility, Rigidity and posture,\\ 
  Advanced non-tremor, Advanced tremor
\end{tabular} & - & Yes  \\
10 & Application user & No & 25m & - & 37 & - & Agglomerative (Complete) & Euclidean & PCA & 3  & 
\begin{tabular}[c]{@{}l@{}}
  Healthy, Better functioning on left, \\ 
  Recovering from drug usage
\end{tabular} & Silhouette & Yes  \\
11 & Application user  & Yes & 28m   & Affected side & 34 & Custom  & Kmeans & Euclidean & PCA & 3 & Rigidity, Bradykinesia, Tremor      & - & No  \\
12 & Application user & Yes & 27m & - & 37 & - & Agglomerative (Complete) & Euclidean & PCA & 5 & 
\begin{tabular}[c]{@{}l@{}}Low Symptoms, Smoothness LV1, Tremor\\ 
  at rest, Smoothness LV2, Medication use
\end{tabular} & -  & No \\  
\bottomrule
\end{tabular}
}
\caption{Results of our user study. Twelve participants, subdivided by data analyst archetype and by domain expertise, were asked to answer the question ``Can you identify phenotypes in Parkinson's disease?'' by analyzing a real-world dataset containing patient data. The table reports the clustering parameters, features adopted and subsets considered in each analysis. Participants were also asked to choose a single clustering instance and assign an interpretable name to the identified clusters.
  \label{table:results}
  \vspace{-1em}
}
\end{table*}

\begin{figure}
\centering
\includegraphics[width=0.5\textwidth]{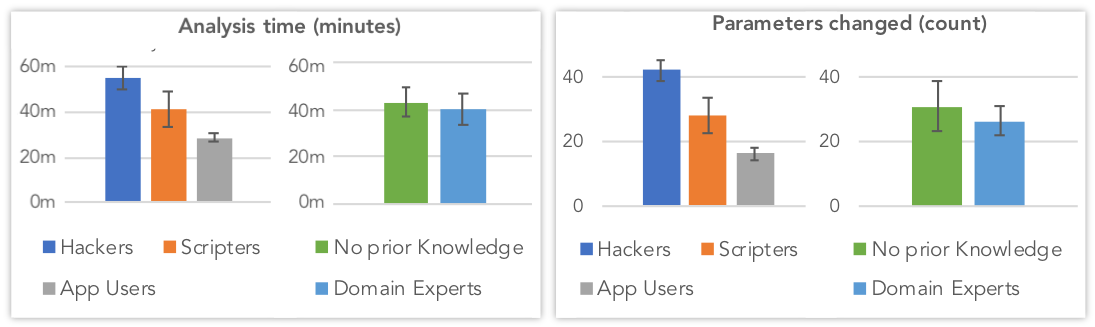}
\caption{Average analysis time and total number of parameters changed, grouped
by data analyst archetype and domain expertise. Hackers seem to be the group
investing more time in the analysis and changing the most parameters (D2).
Domain expertise appears to slightly reduce the average analysis time.
\label{fig:time_parameters}}
\vspace{-0.5em}
\end{figure}

\begin{figure}
\centering
\includegraphics[width=0.5\textwidth]{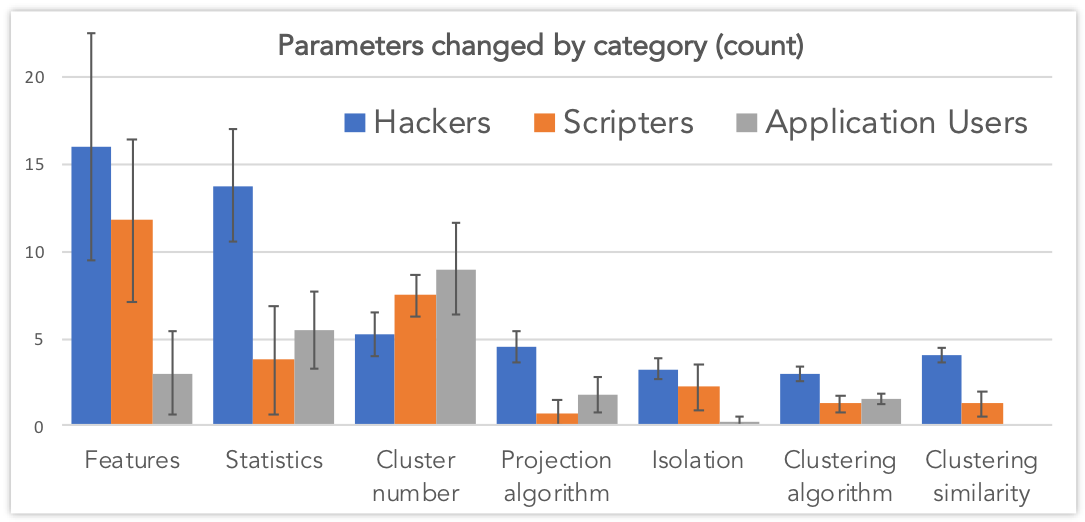}
\caption{Parameters changed during the analysis grouped by category and user
archetype. The bar chart shows that choosing input features (enabling /
disabling data dimensions) was the most performed activity, while clustering
method and metric were changed less often. The cluster number parameter was
also frequently adjusted, especially by the application user archetype in the
context of Agglomerative clustering. It is interesting to note that about 27\%
of the total parameters were repeated by the same user, demonstrating the
highly iterative nature of exploratory data analysis and the need for caching
results (D8). Despite that ``Statistics'' is not a clustering parameter, we report under that name the number of times each user accessed feature distribution
information from the Data Table.
\label{fig:parameters_changed}
\vspace{-0em}
}
\end{figure}

\section{User Study}
We conducted a study with twelve data scientists using \ctwo to answer an open
analysis question about a real-world dataset. We had two goals: 1)
understanding how data scientists might use the interactions, visualizations,
and user-guidance features of our tool based on their level of expertise and
prior knowledge of the data domain, 2) studying the overall workflows adopted
by data scientists to arrive to a solution they consider satisfactory in an
open-ended analysis task about a real-world dataset, where finding a solution
is not guaranteed.  


\bfparhead{Data} 
We chose a real-world dataset concerning subjects with Parkinson's disease, in which
there is not trivial solution to the clustering problem.  The dataset has 8652
rows and 37 features, obtained after preprocessing a subset of the data made
publicly available by the Parkinson’s Progression Markers Initiative (PPMI).
The data contains records about human subjects associated with the Unified
Parkinson's Disease Rating Scale (UPDRS), which consists of a set of measures
that describe the progression of an individual’s condition. The measures are evaluated by interview and clinical observation of human
subjects by a clinician, and include symptoms such as rigidity of upper and
lower limbs, leg agility, gait, spontaneity of movement, finger and toe
tapping, tremor (see ~\cite{goetz2008movement} for the full list of measures). While most features were UPDRS values ranging from 0 to 4, a few
others indicated the overall progression of the disease (Hoen \& Yahr stage),
the use of medication (\textsc{ON\textunderscore OFF\textunderscore DOSE}) and the
number of hours passed from when the subject took a drug 
(\textsc{PD\textunderscore MED\textunderscore USE}). 
\bfparhead{Task}
Participants were asked to complete a single task: ``Identify the different
phenotypes characterizing Parkinson's disease in the given dataset.'' We
defined ``phenotypes'' as the observable behaviors of a subject due to the
interaction of the disease with the environment. We asked our participants to
identify one clustering instance that they were satisfied with, assign a name
and a description to each of its clusters, and finally explain verbally the
significance of their obtained results. 

\bfparhead{Participants} We recruited twelve participants who had worked as data scientists for at least two years.  They all had at least a
master’s degree in science or engineering. To recruit our participants, we
first interviewed 16 candidates and then selected twelve by matching
candidates with the three analyst archetypes~\cite{kandel2012enterprise},
\textit{hackers}, \textit{scripters}, and \textit{application users}, based on
their expertise and domain knowledge. We ensured that we had four participants
for each of the three analyst types. Note that hackers have solid
scripting and programming skills in different languages, such as C++ and
Python, and are capable of developing their own tools for data analysis;
scripters are more familiar with scripting (e.g., using R, Matlab,
etc.) than programming and generally have a robust background in Mathematics or
Statistics; and application users conduct their analysis using
spreadsheet applications such as Microsoft Excel or other off-the-shelf data
analysis tools such as SAS and SPSS.  For each of these three archetypes, we
also made sure that we had  two participants with domain expertise in Parkinson's disease or
neuroscience, and two participants with no prior knowledge about this data domain---for a total of six domain experts and six novices. Participants ranged from 28 to 47 years old, with an even gender distribution.


\bfparhead{Procedure} The study took place in the experimenter’s office, where
one participant at a time used \ctwo on the experimenter’s laptop.
Participants were first briefed on the study and then given a tutorial on
\ctwo  for about fifteen minutes, using the OECD dataset as sample data.  After
the tutorial, participants were  introduced to the test dataset, and the
experimenter explained the medical terminology found in feature names (e.g.
``pronation-supination left hand'').  Regardless of their knowledge of
Parkinson's disease, all participants were tasked with identifying
groups of patients with different phenotypes in the dataset. Participants were
given two hours to provide a ``solution'' with which they were satisfied.  Using
\ctwo's logging feature,  we timestamped and recorded each operation performed
by participants.  During the analysis session, participants were asked to think
aloud and comment on the reasons behind their choices, which we recorded using
an audio recorder.  Participants could conclude the session and stop the timer
whenever they felt they had obtained a satisfactory result.
At the end of the analysis session, participants  were asked to verbally
describe the clusters in their solution, based on insights
derived from their analysis. They also completed a followup questionnaire, 
where they were asked to answer the following questions using free text: 
Q1) ``Are you satisfied with the results or insights you obtained?'',
Q2) ``Would you be able to obtain a better result with another tool or your
own coding skills?'', Q3) ``Did naming clusters help you reason about their
significance?'', and Q4) ``Did \ctwo help you in deciding the clustering parameters?''.

\subsection{Results} 
We summarize the results of our study in Table~\ref{table:results},
reporting for each user their archetype and their expertise in the domain.
The clusters identified by our participants are in line with recent work on
phenotypes in Parkinson's disease~\cite{goetz2008movement,mazzoni2012motor}.
Below we discuss the results of this user study in depth, where we use the
notation P\# to refer to participant number \#, and refer back to our design criteria to validate our design choices. 
We also provide additional insights from the user study in \textit{Supplemental Material}.

\bfparhead{User Archetypes and Domain Knowledge} 
Our study confirmed the relationship between the analyst types 
and attitude in performing data analysis~\cite{kandel2012enterprise}. 
For instance, Fig.~\ref{fig:time_parameters} shows that, on average, hackers  
seem to invest more time trying out more parameter
combinations than the other data analyst types. Similarly, expertise in the
neuroscience domain suggests shorter analysis time, possibly due to
better knowledge of the data features. 
In Fig.~\ref{fig:parameters_changed} we break down the interactive 
parameters changed by participants during their analysis sessions 
into sub-categories (e.g. how many times they enabled/disabled
a feature, how many times they changed the number of clusters). We found that
different archetypes tended to use the features available in
\ctwo differently. Even the type of algorithms and methods used seem to be
correlated to analyst archetypes, as shown in Table~\ref{table:results}.
Overall, participants' answers to Q1 and Q2 demonstrate that \ctwo 
supports the analysis style characteristic to all types of data analysts.

\bfparhead{Analysis Flow} 
For all participants, the analysis started with a default Clustering View
automatically applying PCA and Agglomerative clustering to the data. The first
action performed by five out of twelve users was to select features of
interest in the data, using either the \textit{Data table} or the \textit{Help
me decide} panel. Most domain experts removed the non-UPDRS features directly
from the \textit{Data table}, whereas participants without prior knowledge
often identified them through the \textit{Help me
decide} panel’s feature selection tab.  Five other participants preferred instead to first try out
different clustering algorithms and numbers of clusters, observing the resulting changes
in the scatterplot and in the heatmap.  These users generally later noticed the
high influence of non-UPDRS features such as \textsc{PD\textunderscore
MED\textunderscore DOSE} and \textsc{Hoen \& Yahr}, primarily thanks to the
heatmap visualization. Then they proceeded in a fashion similar to the
domain experts, excluding these features from their subsequent analysis. 
Finally, two out of twelve users (P6 and P9) preferred to start their analysis with the Clustering Tour. In
most cases, the analysis continued with an iterative and cyclic modification of
clustering parameters and selected features, until participants realized that
they could only find clustering outcomes based on affected side or severity of
the disease. These clusters were easily interpreted from the heatmap
visualization, which showed a horizontal gradient for increasing severity
and an alternate pattern in rows corresponding to the left or right side of the patient's body.

Here, some participants made stronger assumptions about the data and applied
different strategies involving subclustering, filtering and feature selection (D3).
In particular, P1, P3 and P9 decided to consider only the features
associated with one part of the body in order to remove the separation in left
and right side-affected subjects. Similarly, P5 and P11 decided instead
to consider only people with one affected side at a time by performing
subclustering. Other participants applied an equivalent strategy by filtering
data points based on affected side and symptom severity, trying to find
relevant insights in a smaller but more significant subset of the original
data.

\subsection{Discussion}
\bfparhead{The Importance of Feature Selection} The first insight we identified
based on the final clustering outcomes in Table~\ref{table:results} and the
histogram in Fig.~\ref{fig:parameters_changed} was the relevance of feature
selection in clustering analysis. More than any other parameter, the choice of
features to feed to the clustering algorithm led users towards
a satisfactory result, and at the same time was the part of the analysis participants spent most of their time on.  In particular, participants
used the feature distribution information available in the Data Table in
combination with the statistical analysis methods available in the ``Help me
decide'' panel (D7).  Whereas domain experts were often able to spot uninteresting
features based on their names (e.g., non-UPDRS features such as \textsc{ON\textunderscore
OFF\textunderscore STATE} and \textsc{PD\textunderscore MED\textunderscore USE}) and
directly remove them from the data table, participants with no prior knowledge
about the domain made heavy use of principal component analysis (PCA) and
univariate feature selection (e.g., ANOVA) to test the relevance of data
dimensions. This allowed scripters in particular to quickly spot features that
were contributing the most to the clustering outcome, and eventually remove
them from the analysis. The hacker archetype often complemented these findings
by inspecting the distribution values (e.g., variance) and pairwise correlations
of each feature from the Data Table. The application users seemed instead
to prefer identifying relevant features and correlations from the horizontal
color distribution of cells in the heatmap, expressing a more qualitative approach.  
After removing a first set of
features, participants generally applied different clustering parameters until
they realized a second round of feature selection was needed.  Here the most
used method was feature agglomeration, with which participants tried to
agglomerate features based on correlation or semantics (e.g. removing features
with high pairwise correlation, keeping only one feature out of four for
tremor, keeping the feature with the highest variance for each left-right
pair). 

\bfparhead{Clustering Tour: Exploring Different Perspectives}
While most participants preferred to adopt trusted parameters, the results in
Table~\ref{table:results} show that the four participants who used the
Clustering Tour were more eager to adopt less conventional algorithms and
metrics, leaving their comfort zone (D7). ``I only pushed a button and it already
gave me insights I would probably not have found by trying out parameter
combinations myself'', commented P9. P4, belonging to the hacker
archetype, stated ``I generally hate automated features that allow script
kiddies to do result shopping (i.e., blindly use system-generated results).
However, \ctwo gives me the possibility to decide myself if solutions
are reasonable. I think it's useful for thinking outside the box.'' 
P9 started using the Clustering Tour in an unconstrained way,
whereas the remaining participants started the tour after first setting 
the number of clusters desired and selecting a subset of the input 
features.  The average number of solutions generated before
a participant expressed positive feedback was 3.7, followed by an average of
2.3 iterations in the solution neighborhood. In particular, the Clustering Tour
proved to be useful in removing non-relevant features and randomly ``shuffling''
data dimensions, generating new perspectives on the analysis. P6, for
instance, performed his analysis without noticing the large bias introduced by
the \textsc{PD\textunderscore MED\textunderscore USE} feature until a solution
generated by the Clustering Tour excluded it from the analysis, showing a
semantically different clustering result. Similarly, P9 realized he could
agglomerate features associated with similar tasks after the Clustering Tour
proposed a solution based on removing highly correlated features.

\bfparhead{(In)Effectiveness of Validation Measures} 
The results of our user study show that validation measures do not perform well
in the presence of specific goal-oriented questions that go beyond pure exploration
of the data.  While most participants did not even consider the use of
validation measures, four participants made use of the ``'Help me decide'' and
the ``Is this a good clustering?'' panels to try to compare measures among
different number of clusters and across clustering instances (D6).  However,
theoretical best cluster separation often suggested considering two or three
clusters, less than what we would generally expect while searching for
phenotypes. In most cases, changing clustering parameters according to
validation measures generally produced clustering outcomes with a different and
possibly less interesting semantic meaning. P12 commented ``I think it makes
more sense to see first if the clusters are interesting from the heatmap, and
then simply check if they are not too overlapping in the scatterplot''.  We
believe validation measures are effective in comparing clustering results only
when the latter are not too semantically different from each other (i.e. low
$\Delta l$). In particular, we can use validation metrics to filter and rank the
solutions automatically generated by our Clustering Tour after the user has
expressed positive feedback (i.e. when most influential parameters have been
fixed). Separately, we can also use them to select the best projection to
visualize a clustering result.

\bfparhead{Cluster Naming and Interpretability} 
According to the answers to Q3, seven out of twelve participants stated that
having to verbally describe and name clusters through the tool interface
significantly helped them better reason about the clustering instance they
found (D5). ``I personally never did it, but giving a name to a cluster forces
you to go beyond the superficial instinct of finding well separated groups. It
often makes you realize you are not searching for the right things'', commented
P5.  Ten participants named their final clusters only by interpreting the
colors of each column in the heatmap, whereas two of them complemented this
information with the automated cluster descriptions in the ``Is this a good
clustering?'' panel (D6). This proves that the heatmap visualization can be a
powerful and self-descriptive yet simple way to semantically summarize
clustering results. Likely related to this, and to the fact cluster members
changed often during the analysis, only one participant used the cluster naming
functionality before being required to provide the final clustering solution.
Naming clusters automatically based on the centroid also did not generalize to
this dataset, where data points were named based on the subject's numerical
identifier.  The automatically generated textual descriptions for clusters that
we introduced in this paper are not fit for systematically assigning short,
meaningful names to clusters. However, a possible solution could be to generate
cluster identifiers semi-automatically by incorporating user feedback on
primary features of interest and how they are semantically related to each
other.


\section{Conclusion}
We present \ctwo, a new interactive tool that guides users in exploratory
clustering analysis, adapts user feedback to improve user guidance, facilitates
the interpretation of clusters, and helps quickly reason about differences
between clusterings. \ctwo introduces the Clustering Tour to assist users in
efficiently navigating the large space of possible clustering instances. We
evaluate \ctwo through a user study with 12 data scientists in exploring
clusters in a dataset of Parkinson's disease patients. 

Our work here confirms that clustering analysis is a nontrivial process, which
requires iterative experimentation over different clustering parameters 
and algorithms as well as data attributes and instances.  
We find that, despite the fact that different users exhibit different attitudes
towards exploratory analysis, feature selection is where they spend most of
their effort.  We also find precomputation to be essential for supporting
interactive analysis. Fig.~\ref{fig:parameters_changed} shows, in fact, that
dynamically changing the number of clusters was a frequent interaction
performed by participants. Given the number of repeated operations, caching
also proved to be essential. While the relevance of user assumptions and prior
knowledge on the data further confirmed that clustering cannot be automated
without incorporating these concerns, participants showed a tendency to stick
to well-known parameter combinations or blindly attempted multiple combinations
by trial and error. This is where the system can come into play and assist the
user in making parameters more explainable or in comparing alternative choices
through the support of statistical analysis. The Clustering Tour we introduce
here demonstrates how to nudge users to think outside the box in exploratory
clustering analysis, avoiding premature fixation on certain attributes or
algorithmic and parametric choices. 

Finally, the ability to compare clusterings through validation metrics does not
necessarily improve the clustering interpretability. To this end, we urge the
development of clustering tools that facilitate explainability, while keeping
in mind that the usefulness of a clustering outcome mostly depends on the
underlying data and user task.

\bibliographystyle{abbrv-doi}
\balance
\bibliography{paper}

\end{document}